\documentclass[spconf,a4paper]{article}
\usepackage{spconf}

\twocolumn
\usepackage{graphicx,color}
\usepackage{amsmath, amsthm, amsfonts, amssymb, amsbsy,nccmath}
\usepackage{mathtools}

\usepackage{algorithm}
\usepackage{enumerate}
\usepackage{lipsum}

\usepackage{textgreek}
\usepackage{textcomp}

\usepackage{algorithmic}
\usepackage[sort,compress]{cite}
\usepackage{epsfig}
\usepackage{epstopdf}
\usepackage{mathtools}
\usepackage{dsfont}
\usepackage{epstopdf}

\usepackage{sidecap, caption}

\theoremstyle{definition}

\theoremstyle{assumption}

\theoremstyle{proposition}

\theoremstyle{corollary}

\usepackage[inline]{enumitem}   
\makeatletter
\newcommand{\inlineitem}[1][]{%
\ifnum\enit@type=\tw@
    {\descriptionlabel{#1}}
  \hspace{\labelsep}%
\else
  \ifnum\enit@type=\z@
       \refstepcounter{\@listctr}\fi
    \quad\@itemlabel\hspace{\labelsep}%
\fi}
\makeatother
\newcommand\norm[1]{\left\lVert#1\right\rVert}

\newtheorem{theorem}{Theorem}

\newcommand{\beq}{\begin{equation}}
\newcommand{\eeq}{\end{equation}}

\newcommand{\p}{{\!\, +\!\,}}

\newcommand{\mV}{\mathcal{V}}
\newcommand{\mW}{\mathcal{W}}
\newcommand{\mZ}{\mathcal{Z}}
\newcommand{\mY}{\mathcal{Y}}

\newcommand{\mX}{{\mathcal X}}

\newcommand{\dsum}{\displaystyle\sum}

\def\adots{\mathinner{\mskip0mu\raise0pt\vbox{\kern7pt\hbox{.}}\mskip3mu
          \raise4pt\hbox{.}\mskip3mu\raise8pt\hbox{.}\mskip0mu}}

\newcommand{\bmW}{{\boldsymbol W}}
\newcommand{\bmV}{{\boldsymbol V}}

\newcommand{\tr}{\mbox{tr}}

\usepackage{bm}

\newcommand{\bmx}{{\bm x}}
\newcommand{\bmy}{{\bm y}}
\newcommand{\bmv}{{\bm v}}

\newcommand{\bmH}{{\bm H}}

\newcommand{\bmR}{{\bm R}}

\newcommand{\bmzh}{\widehat{\bmz}}

\newcommand{\bmS}{{\bm S}}

\newcommand{\bmA}{{\bm A}}
\newcommand{\bmM}{{\bm M}}

\newcommand{\bmHh}{\widehat{\bmH}}

\newcommand{\bmI}{{\boldsymbol {I}}}

\newcommand{\bma}{{\bm a}}

\newcommand{\btheta}{\boldsymbol{\theta}}
\newcommand{\balpha}{\boldsymbol{\alpha}}
\newcommand{\bbeta}{\boldsymbol{\beta}}

\makeatletter
\newcommand\fs@spaceruled{\def\@fs@cfont{\bfseries}\let\@fs@capt\floatc@ruled
  \def\@fs@pre{\vspace{0.5\baselineskip}\hrule height.8pt depth0pt \kern2pt}%
  \def\@fs@post{\kern1pt\hrule\relax}%
  \def\@fs@mid{\kern2pt\hrule\kern2pt}%
  \let\@fs@iftopcapt\iftrue}
\makeatother

\newcommand{\E}{\mbox{E}}

\newcommand{\bit}{\begin{itemize}}
\newcommand{\eit}{\end{itemize}}

\newcommand{\mG}{\mathcal{G}}

\newcommand{\mI}{\mathcal{I}}

\newcommand{\bmz}{\boldsymbol{z}}
\newcommand{\zh}{\widehat{z}}

\newcommand{\kbar}{\overline{k}}

\newcommand{\bmVh}{\widehat{\bmV}}

\newcommand{\bphi}{\boldsymbol{\phi}}

\newcommand{\ibar}{\overline{i}}
\renewcommand{\bmA}{{\boldsymbol A}}

\newcommand{\bmX}{{\boldsymbol{X}}}

\newcommand{\bmB}{{\boldsymbol B}}

\newcommand{\bmN}{{\boldsymbol N}}

\DeclarePairedDelimiter\abs{\lvert}{\rvert}%

\renewcommand{\bmS}{{\boldsymbol S}}

\newcommand{\bmZ}{{\boldsymbol Z}}

\newcommand{\bmzt}{\widetilde{\bmz}}

\newcommand{\bmZh}{\widehat{\bmZ}}

\newcommand{\bmHt}{\widetilde{\bmH}}

\newcommand{\bmWh}{{\widehat{\bmW}}}
\newcommand{\bmVt}{{\widetilde{\bmV}}}
\newcommand{\bmzhh}{{\doublehat{\bmz}}}
\usepackage{lipsum}
\usepackage{accents}
\newlength{\dhatheight}
\newcommand{\doublehat}[1]{%
    \settoheight{\dhatheight}{\ensuremath{\hat{#1}}}%
    \addtolength{\dhatheight}{-0.35ex}%
    \hat{\vphantom{\rule{1pt}{\dhatheight}}%
    \smash{\hat{#1}}}}

\restylefloat{algorithm}
\usepackage{geometry}
\begin{document}
\ninept
\title{ Reliable Beamforming at Terahertz Bands: Are Causal Representations the Way Forward? \vspace{-4.5mm}}
\vspace{-8mm}
\name{Christo Kurisummoottil Thomas and Walid Saad\thanks{This research was supported by the Office of Naval Research (ONR) under MURI grant N00014-19-1-2621.}\vspace{-4mm}}
\vspace{-20mm}\address{Wireless@VT, Bradley Department of Electrical and Computer Engineering, \\ Virginia Tech, Arlington, VA, USA, \\ \fontsize{6}{7}Emails: \{christokt,walids\}@vt.edu\vspace{-8mm}}
\maketitle
\vspace{-6mm}
\begin{abstract}\vspace{-2mm}
Future wireless services, such as the metaverse require high information rate, reliability, and low latency. Multi-user wireless systems can meet such requirements by utilizing the abundant terahertz bandwidth with a massive number of antennas, creating narrow beamforming solutions. However, existing solutions lack proper modeling of channel dynamics, resulting in inaccurate beamforming solutions in high-mobility scenarios. Herein, a dynamic, semantically aware beamforming solution is proposed for the first time, utilizing novel artificial intelligence algorithms in variational causal inference to compute the time-varying dynamics of the causal representation of multi-modal data and the beamforming. Simulations show that the proposed causality-guided approach for Terahertz (THz) beamforming outperforms classical MIMO beamforming techniques.

\end{abstract}

\vspace{-5mm}
\section{Introduction}
\vspace{-3mm}
Future wireless systems must communicate massive multi-modal sensory information to enable emerging applications such as the metaverse and extended reality (XR) \cite{ChaccourITJ2022,YangAccess2018,OmarICC2023}. However, sub-6 GHz and millimeter wave (mmWave) bands have limited bandwidth and cannot satisfy the stringent quality-of-service (QoS) requirements of XR applications in terms of delivering high data rates, low latency, and high reliability. Integrating XR services over high-frequency terahertz (THz) bands is a promising solution, but the wireless channel at THz frequencies is highly susceptible to significant blockage effects, limiting the significant multipath components. As such, pencil-like narrow beamforming (BF) solutions that can be dynamically adjusted are required to provide seamless connectivity for users, especially in high mobility scenarios.

\begin{figure}[t]
\centering{\includegraphics[width=3.1in,height=1.4in]{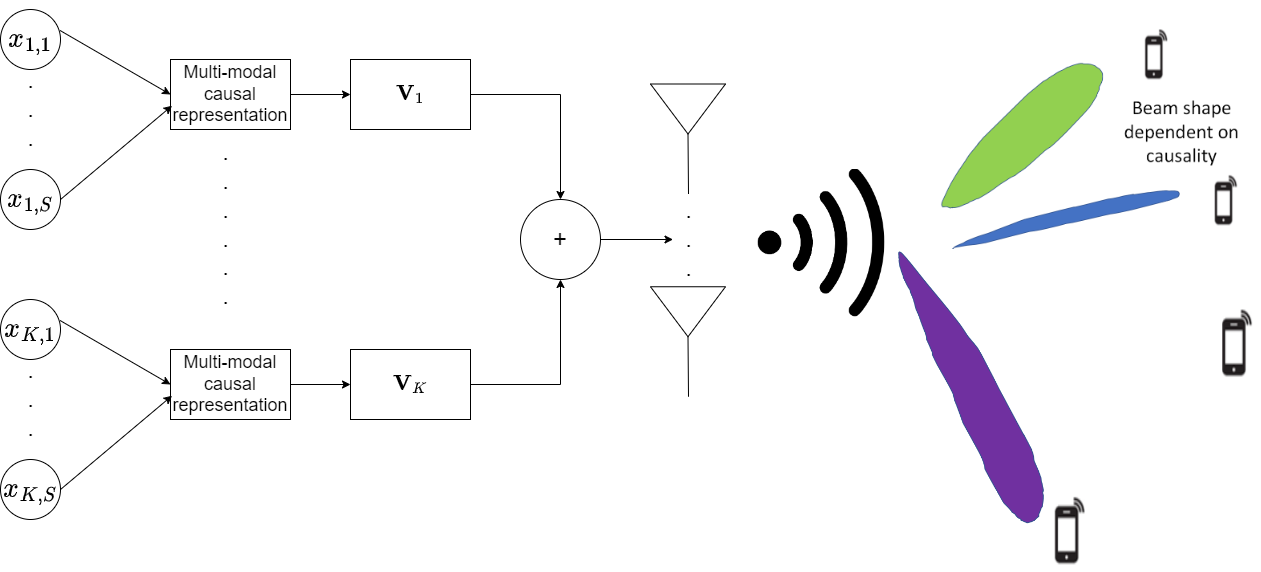}}\vspace{-3mm}
\caption{Different BFs can have variable beam width or shape based on the semantic richness of the transmitted content. }
\label{CausalityAwareBF}\vspace{-0mm}
\vspace{-6mm}
\end{figure}

To enable pencil-like BF solutions, usage of (ultra)-massive multiple-input multiple-output (MIMO) antenna systems (which becomes practical due to smaller antenna spacing) is expected \cite{ChaccourCST2022}. However, accurately tracking time-varying user channels and mitigating downlink interference caused by inaccurate BF solutions (i.e., leakage from sidelobes) to other users in the network are critical challenges in providing high-rate communication links using BF. Conventional approaches here \cite{Al-TanDaiJSAC2021} is to either use a codebook-based beam tracking method or channel information-based BF by exploiting Kalman filtering (KF)-based methods \cite{Al-NaffouriTSP2007} to track the time-varying channels while considering a specific user mobility model. The first method suffers from a large codebook size (hence, larger overhead) to follow a narrow beam direction along the user, particularly at THz bands. The second scheme need not be practical due to the user mobility model, which may be inaccurate and
cannot be easily modeled. Another major direction is utilizing artificial intelligence (AI) based algorithms as in \cite{MashhadiTWC2010}. However, those techniques require a larger amount of data and higher training overheads. A promising candidate here that has not been explored yet is to design AI-native wireless systems \cite{ChaccourArxiv2022} that can extract the causal aspects (why and how the data gets generated) present in the wireless environment and the content to be transmitted. Such causal information would represent the semantics of the channel and data. We envision that transmitting just the semantic substance instead of the irrelevant components present in the data has two benefits. Firstly, it reduces the number of physical bits to be transmitted, thus improving communication resource efficiency \cite{PopovskiJIIS2020,StrinatiComNetworks2021,KountourisCommMag2021,XieTSP2021}. Second, in a THz system, we envision that a causality-aware BF solution could improve reliability at the end user due to increased BF gain. This means that a semantic-aware system can judiciously choose the null space of the BF vectors, thus leaving enough dimensions to increase the gain along the specific users that have semantically rich data. However, none of the existing works in the literature deal with multi-user semantics except \cite{HuiqiangJSAC2022}. However, \cite{HuiqiangJSAC2022} limits the discussions to the design of the encoding architecture while relying on a conventional BF approach.  Contrary to \cite{HuiqiangJSAC2022}, for the design of multi-user semantic communications in a THz system for a metaverse application, we try to address the following questions:
\vspace{-1mm}\begin{itemize}
\item What is the best way to represent multiple modalities at the base station (BS) in a shared semantic subspace that can capture causal reasoning across modalities while preserving the modality-specific causal aspects?\vspace{-2mm}
\item How to design a BF vector for each user in a THz system,  guided by the causality dynamics across users so that it can mitigate the impact of multi-user interference and simultaneously preserve the causality aspects of the multi-modal data while ensuring that the users can decode it reliably?
\end{itemize}\vspace{-1mm}
The main contribution of the paper is, thus, a novel multi-user, multi-modal communication system that can deliver reliable THz transmission by exploiting the causality dynamics present in the channel and source data. In particular, we propose a novel BF solution that can be represented as a linear combination of a static component and a dynamic component. The static part, is a nonlinear function of the causality dynamics of the channel and user data and learned using variational causal networks (VCN) \cite{AnsonArxiv2022}. The dynamic component must be adjusted based on instantaneous channel estimates and is solved by maximizing the minimum semantic information across all users. The resulting BF solution is semantics (causality) aware, see for e.g., Figure.~\ref{CausalityAwareBF}. This means that the subspace dimension over which the BF must mitigate interference varies depending on the semantic richness of information at interfering UEs. Compared to classical AI schemes that rely on uncertainty or statistics present in the data, exploiting the causal aspects enables the system to learn with dramatically less data and reduces the training overhead \cite{ChristoTWCArxiv2022}. Our proposed framework
achieves a high semantic information rate and semantic reliability compared to conventional schemes that use beam tracking or uplink channel estimation-based beamforming solutions (without proper Doppler modeling). 

\vspace{-5mm}
\section{System Model}
\vspace{-4mm}

Consider a wireless network in which a BS transmits i.i.d multimodal metaverse content in the downlink (DL) to a set $\mathcal{K}$ of $K$ user equipment (UE) over THz bands. The BS has $M$ antennas, and each UE has $N$ antennas. The data to be transmitted to user $k$, is $\bmx_k = \left[x_{k,1},\cdots,x_{k,S}\right]$, with $x_{k,m}$ corresponding to modality $m$. Modality here refers to data from a separate source. Transmitting this raw multi-modal data directly to users is delay-critical and requires a high data rate. Instead, we propose to compute a latent representation that can explain the causes behind the data generation, called \emph{causal representation}. The causality (that represents the \emph{semantics} here) of the multi-modal data for UE $k$, is defined as $\bmz_k \in \mathcal{R}^D$. We assume that $\min(M,N) \geq D$. The linear BF matrix $\bmV_k$ for user $k$ serves to project the semantics $\bmz_k$ in a subspace ($\subset \mathcal{R}^M$) such that the UE can reconstruct the received metaverse content with high semantic reliability. \emph{Semantic reliability} here is defined as the accuracy of the semantics reconstructed at the
receiver compared to the intended semantics transmitted from the BS. We chose to represent the BF and causality of the data using two components, motivated by the formulations in KF \cite{RibeiroISR2004} under nonlinear state dynamics, with the $\bmVt_k, \bmzt_k$ representing the prediction from the unknown nonlinear dynamics of user channels and causality. The BF matrix, at instance $t$, of dimensions $M\times D$ is $\bmV_k^{(t)} = \lambda\bmVh_k^{(t)} + (1-\lambda)\bmVt_k^{(t)}$, where $\bmVt_k^{(t)}$ represents the component that is a function of the channel statistics (derived from the channel history) and semantics $\bmz_k$ until time $t-1$ and $\bmVh_k^{(t)}$ represents an instantaneous component of the BF vector that tracks dynamic variations in the channel and causality. $\lambda \in [0,1]$ is a constant factor. Similarly, the causal representation is also assumed to have a linear representation whereby $\bmz_k^{(t)} = \beta\bmzh_k^{(t)} + (1-\beta)\bmzt_k^{(t)}$, with $\beta \in [0,1]$ being a constant factor. The receive combiner $\bmW_k$ is a matrix of dimension $N\times D$ and is assumed to only have an instantaneous component. A semantic-aware receive combiner design follows similarly but is not discussed due to space limitations. We can write the received signal at user $k$ after the combining as follows (index $t$ is omitted for notational convenience):
\vspace{-2mm}\beq\vspace{-1mm}
\bmy_k = \bmW_k^H\bmH_k\sqrt{p_k}\bmV_k\bmz_k + \bmW_k^H\bmH_k\sum\limits_{i\neq k}\sqrt{p_i}\bmV_i\bmz_i + \bmW_k^H\bmv_k. \label{eq_obs_model}
\vspace{-3mm}
\eeq
Here, $p_k$ is UE $k$'s transmit power, assumed to be fixed and equal for all users. The entries of noise vector $\bmv_k$ follow $\mathcal{N}(0,1)$. 
The THz channel at multi-path delay $d$ between any user and the BS is \cite{PriebeTWC2013}:
\vspace{-3mm}\beq
\begin{array}{l}
\bmH_d = \alpha_0G_tG_rp_r(dT_s-\tau_0)e^{j2\pi\nu_0t}\bma_r(\theta_0)\bma_t(\phi_0)^T + \\ \sum\limits_{i=1}^P\alpha_iG_tG_rp_r(dT_s-\tau_i)e^{j2\pi\nu_it}\bma_r(\theta_i)\bma_t(\phi_i)^T,
\end{array} \label{eq_channel_model}
\vspace{-3mm}\eeq
where $\alpha_i$ represents the complex path gain, $\theta_i,\phi_i$ are the AoA and AoD, respectively, and $G_t,G_r$ represents the antenna gains at BS and UE, respectively. $\nu_i$ is the Doppler frequency corresponding to path $i$.  Parameters with subscript $0$ correspond to line-of-sight (LOS) components. The number of non-LOS (NLOS) paths $P$ is considered small as is typical in THz \cite{ChaccourITJ2022}. The LOS channel gain will be \cite{ChaccourICC2022} $\alpha_0 = \frac{c}{4\pi f r}e^{-\frac{\kappa(f)r}{2}}e^{-j2\pi f\tau_0}$, where $\kappa(f)$ is the overall molecular absorption coefficient of the
medium at THz band, $f$ is the operating frequency, $c$ is the speed of light, and $r$ is the distance between the user and the BS. Further, we can write the channel at any subcarrier $f$ as
$\bmH_f = \sum\limits_{d=0}^{N_p-1}\bmH_d e^{\frac{-j2\pi f d}{N_s}}.$
However, we consider the BF design for a single subcarrier for simplicity. Hence, hereinafter, we ignore the notation $f$. This is also motivated by the fact that the semantics transmitted across different subcarriers are independent, and hence, we can consider the BF design separately for each subcarrier. 

We now briefly touch upon the channel estimation characterization that motivates the BF and $\bmz_k$. Consider that the BS has the linear minimum mean squared error (LMMSE) channel estimates computed using uplink (UL) pilots, whose behavior can be characterized as
$
 \bmH_k = \bmHh_k + \bmHt_k,$
  where the estimates $\bmHh_k \perp \bmHt_k$ lie in orthogonal subspace. $\bmHt_k$ includes both the white noise component in the received pilots and the time-varying part (assuming DL transmission slots differ from the UL pilots). Motivated by the above formulations for the channel estimates, we split the problem into two parts: a) the time-varying component ($\bmVt_k, \bmzt_k$) computed using an AI approach and is a function of the channel dynamics, thus a function of the statistics $\mathbb{E}(\bmHt_k\bmHt_k^H)$, and b) the instantaneous tracking part ($\bmVh_k, \bmzh_k$) using $\bmHh_k$ formulated as a non-convex optimization, to maximize the minimum semantic information. Specifically, we propose to learn the time-varying dynamics (computed as the posterior distribution) of the BF and causal components from the channel estimation history. Components $\bmVt_k$ and $\bmzt_k$ are learned as the mode of the posterior distribution:
  \vspace{-1mm} \beq
   \vspace{-2mm}[\bmVt_k^t, \bmzt_k^t] = \arg\max\limits_{\bmVt_k,\bmzt_k} p(\bmz_k,\bmV_k\mid \mathcal{H}^{0:t-1},\mZ^{0:t-1}),\label{eq_mode_pos}
  \vspace{-1mm} \eeq
 where $\mathcal{H}^{(0:t-1)}$ and $\mZ^{0:t-1}$ are the set of all channel matrices and causal representations till time $t\!-\!1$. Next, we look at the problem formulation to compute the BF matrices, combiners and $\bmz_k$. 

\vspace{-5mm}\section{Problem Formulation}
\vspace{-3mm}

Given instantaneous (and perfect) channel information, our objective is to compute the BF/combiner matrices $\bmV_k$ and $\bmW_k$  as well as the causal representation $\bmz_k$ such that the semantic reliability at the UE $k$ is above a determined threshold. We denote $\bmzhh_k$ as the reconstructed semantics at the UE. We consider max-min average semantic information as the optimization criteria so as to ensure fairness with respect to user rates. We now introduce relevant metrics, whose detailed definitions  appear in \cite{ChristoTWCArxiv2022}. 
The channel imperfections result in a \emph{semantic distortion}, which can be captured by the Frobenius norm of the error in prediction by using the transmitted state ($\bmz_k$) and the user's extracted state ($\doublehat{\bmz}_k$): $E(\bmz_k,{\doublehat{\bmz}_k}) = \norm{\bmz_k-\doublehat{\bmz}_k}^2.$
We also define the error measure $E({S},{\widehat{S}}) = \abs{S-\widehat{S}}^2$ in terms of the difference in semantic information ($S$) conveyed by the transmitter and that learned by the receiver.
These semantic distortion measures help to quantify how much semantic information the receiver can extract from a corrupted version of the message. The semantic message space (representing semantic similarity) corresponding to $\doublehat{\bmz}_k$ can be described as 
$
E(\bmz_k,{\doublehat{\bmz}}_k)  \leq \delta,\,\, \mbox{s.t.}\,\, E({S}(\bmz_k),{\widehat{S}}(\doublehat{\bmz}_k;\bmy_k)) = 0.$
$\delta$ represents the threshold in the Euclidean space between two $D$-dimensional causal vectors $\bmz_k,\doublehat{\bmz}_k$ within which the received semantic information is the same (\emph{semantic space} corresponding to $\bmz_k$). As long as decoded messages $\bmzhh_k$ are within the semantic space, it is successfully decoded; hence the \emph{semantic reliability} can be captured by the probability of successful transmission $p(E(\bmz_k,{\doublehat{\bmz}}_k)  \leq \delta)$.
Using these metrics, we formulate our problem (for communication instance $t$) as:
\vspace{-1mm}\beq
\vspace{-1mm}\begin{array}{l}
\mathcal{P}_1: \max\limits_{\mV,\mW,\mZ} \min\limits_k \mathbb{E}_{q}[S_k(\bmz_k;\bmy_k)] \\
\hspace{0.7cm}\mbox{subject\,to}\,\,\,p\left(E({\bmz}_k,{\doublehat{\bmz}}_k) < \delta \right) \geq 1-\epsilon,
\end{array}
\label{eq_P1}
\eeq
where $\mV,\mW, \mZ, \mathcal{H}, \mY,\mX$ represent the set of all $\bmV_k,\bmW_k, \bmz_k, \bmHh_k$, $\bmy_k$ and $\bmx_k$, respectively, and $\epsilon$ is close to $0$. Here, the expectation is over $p(\bmz_k,\bmV_k\!\mid\!\mathcal{H}^{0:t-1},\mZ^{0:t-1})$ and represents the (learned) statistics over channel and data history (see Section~\ref{Section_VCN}). Moreover, our problem formulation relies on novel semantic information metrics, unlike state of the art semantic communication systems that exploit classical information theoretic schemes \cite{Carnap52}. 
Next, we define the semantic information measure required to compute the objective $\mathcal{P}_1$. 
 
\vspace{-5mm}\subsection{Semantic Information Measure}
\vspace{-2mm}

Using category theory, we can define the concept of semantic information, based on \cite{ChristoTWCArxiv2022}, as follows. 
The extracted semantic information at the receiver can be written as \eqref{eq_semInfo_listener}, derived in \cite{ChristoTWCArxiv2022}:
\begin{figure*}\beq
\vspace{-0mm}\begin{array}{l}
\mathbb{E}_{q}\left[S_k(\doublehat{\bmz}_k;\bmy_k\mid \bmz_k,\bmH) \right]   = \sum\limits_{\bmz_k}p\left(\bmy_k\mid \bmz_k\right)S_s({\bmz}_k)\left[\sum\limits_{\doublehat{\bmz}_k}p(\doublehat{\bmz}_k\mid \bmy_k,[\bmzhh_{t-1}])\log\frac{p(\doublehat{\bmz}_k\mid \bmy_k,[\bmzhh_{t-1}])}{p(\doublehat{\bmz}_k\mid[\bmzhh_{t-1}])}Z_{\doublehat{z}_kz_k}\right] 
\\  \stackrel{(a)}\leq\sum\limits_{\bmz_k}p\left(\bmy_k\mid \bmz_k\right)S_s({\bmz}_k)[\sum\limits_{\doublehat{\bmz}_t}\log\det(\bmR_{\kbar}^{-1}\bmR_k)Z_{\doublehat{z}_kz_k}] ,
\end{array}
\label{eq_semInfo_listener}
\vspace{-6mm}
\eeq\vspace{-4mm}
\end{figure*}
with $Z_{\doublehat{z}_kz_k}$ being the similarity between transmitted and extracted causal states as also defined in \cite{ChristoTWCArxiv2022}. $\bmR_{\kbar} = \sum\limits_{i\neq k}\mathbb{E}_q(\bmW_k^H\bmHh_k\bmV_i\bmz_i\bmz_i^T\bmV_i^H\bmHh_k^H\bmW_k) + \sigma^2\bmI$ is the interference-plus-noise covariance matrix at UE $k$. The signal-plus-interference-plus-noise covariance matrix at UE $k$ gets represented as $\bmR_{k} =   \mathbb{E}_q(\bmW_k^H\bmHh_k\bmV_k\bmz_k\bmz_k^T\bmV_k^H\bmHh_k^H\bmW_k) + \bmR_{\kbar}$. In \eqref{eq_semInfo_listener}, (a) follows from first considering that $\bmR_{\kbar}$ converges to its expectation due to the law of large numbers (in the limit $DK\rightarrow \infty$). Further, we can take the expectation operator inside the log using Jensen's inequality, resulting in the upper bound to the semantic information measure. We can expand the expectation (for a fixed $\bmz_k,\bmW_k$) as follows:
\vspace{-2mm}
\beq
\vspace{-2mm}
\begin{array}{l}
\!\!\!\mathbb{E}_q\left(\bmW_k^H\bmHh_k\bmV_i\bmz_i\bmz_i^T\bmHh_k^H\bmV_i^T\bmW_i\right)\! =\! \bmW_k^H\bmHh_k\left(\lambda^2\E_q\left(\bmVt_i\bmz_i\bmz_i^T\bmVt_i^H\right)\right. \\ \left. \!\!\!\!\!\!+  \lambda^2\tr\{\bmz_i\bmz_i^T\}\E_q(\bmVt_i\bmVt_i^H) +  (1-\lambda)^2\bmVh_i^H\bmz_i\bmz_i^T\bmVh_i \right) \bmH_k^H\bmW_k + f(\bmVh_i).
\end{array}\label{eq_Exp}
\vspace{1mm}
\eeq
Here, we assume that the mean $\mathbb{E}_q(\bmVt_{i})$ is zero. $f(\bmVh_i)$ represents the quadratic terms that depends only on $\bmVh_i$. To compute
the expectations in \eqref{eq_Exp}, we need to know the posterior given the history
of observations and channel, which we propose to approximate
using variational causal networks as discussed in Section~\ref{Section_VCN}.

\vspace{-4mm}
\subsection{Causality Aware BF with Generalized Eigen Vectors (GEV) }
\vspace{-2mm}

Since the optimization problem in \eqref{eq_P1} is non-convex with respect to joint variables $\{\bmVh_k,\bmWh_k,\bmz_k\}$, we adopt the standard technique of alternating optimization. We propose to solve $\mathcal{P}_1$ using two sub problems. The BF and combining vectors (for any given $\bmz_k$) can be alternatively computed by solving the following problem.
\vspace{-3mm}\beq
\vspace{-1mm}\begin{array}{l}
\mathcal{P}_2:\!\max\limits_{\bmV_k,\bmW_k}\underbrace{\sum\limits_{\doublehat{\bmz}_k}w_{\zh_k}\log\det(\bmR_{\kbar}^{-1}\bmR_k)}_{\mbox{Concave part w.r.t } \bmV_k} + \\ \underbrace{\sum\limits_{i\neq k}\sum\limits_{\doublehat{\bmz}_k}w_{\zh_k}\log\det(\bmR_{\ibar}^{-1}\bmR_i)}_{\mbox{Convex part w.r.t }\bmV_k}
\end{array}
\label{eq_P2}
\vspace{-2mm}
\eeq
where $w_{\zh_k}=p\left(\bmy_k\mid z_k\right)S_s({z}_k)Z_{\doublehat{z}_kz_k}$. \eqref{eq_P2} is non-concave since it is a summation of concave and convex functions. Hence, this can be solved by constructing an approximate function that is a lower bound to the \eqref{eq_P2}. Next, we derive the BF and combiner matrices by alternating optimization of the resulting approximate function. 
\vspace{-1mm}\begin{theorem}
\vspace{-1mm}\label{theorem_gev}
The BF and combiner matrices corresponding to any user $k$ can be obtained as a GEV of two matrices representing a compromise between maximizing the useful signal power part and minimizing the leakage power part. The corresponding expressions can be wrriten as follows: 
\vspace{-3mm}\beq
\begin{array}{l}
\textrm{vec}(\bmVh_k) = \textrm{G.E.V}(\bmS_{t,k},\bmI_{t,k}), \,\,
\bmW_k = \textrm{G.E.V}_{1:D}(\bmS_{r,k},\bmI_{r,k}), \,\,\\
\mbox{where}, \,\, \bmS_{t_k} = \left(\bmz_k\bmz_k^H \otimes \bmHh_k^H\bmR_k^{-1}\bmHh_k\right), \\ \bmS_{r,k} = \bmHh_k\bmV_k\bmz_k\bmz_k^H\bmV_k^H\bmHh_k^H, \\ \bmI_{t,k} = \dsum\limits_{i\neq k} \dsum\limits_{\widehat{z}_i}w_{\zh_i}\left(\bmz_k\bmz_k^H \otimes \bmHh_i^H(\bmR_{\ibar}^{-1}-\bmR_{i}^{-1})\bmHh_i\right), \\
\bmI_{r,k} = \dsum\limits_{i\neq k} \dsum\limits_{\widehat{z}_k}w_{\zh_k}\bmHh_k\bmV_i\bmz_i\bmz_i^H\bmV_i^H\bmHh_k^H,
\vspace{-3mm}\end{array}
\label{eq_gev_theorem}
\vspace{-1mm}\eeq
where $\textrm{vec}(\bmX)$ represents the vectorized version (by stacking column by column) of the matrix $\bmX$.
\vspace{-2.5mm}
\end{theorem}
\emph{Proof}: To obtain this, we can linearize the non-concave part using a first order Taylor series expansion. This follows similar approach called difference of convex functions as in \cite{GiannakisTIT2011}, hence we skip detailed derivations and convergence analysis. This leads to:
\vspace{-1mm}\beq
\vspace{-2mm}\begin{array}{l}
\max\limits_{\bmV_k}\sum\limits_{\widehat{z}_k}w_{\zh_k} \log\det(\bmR_{\kbar}^{-1}\bmR_k) - \\ \sum\limits_{i\neq k}\sum\limits_{\widehat{z}_i}w_{\zh_i}\tr\{\bmHh_i^H(\bmR_{\ibar}^{-1}-\bmR_i^{-1})\bmHh_i\bmV_k\bmz_k\bmz_k^H\bmV_k^H\}.
\end{array}
\label{eq_V_opt_approx}
\vspace{-1mm}\eeq
Taking derivative of the \eqref{eq_V_opt_approx} w.r.t $\bmV_k$ leads to the following generalized eigen vector (G.E.V) solution. Here, we make use of the relation $\textrm{vec}(\bmA\bmX\bmB) = (\bmB^T\otimes \bmA)\textrm{vec}(\bmX)$.
\vspace{-1mm}\beq
\vspace{-2mm}\begin{array}{l}
\sum\limits_{\bmzh_k}w_{\zh_k}\left(\bmz_k\bmz_k^H \otimes \bmHh_k^H\bmR_k^{-1}\bmHh_k\right)\textrm{vec}(\bmV_k)  = \\ \dsum\limits_{i\neq k} \dsum\limits_{\widehat{z}_i}w_{\zh_i}\left(\bmz_k\bmz_k^H \otimes \bmHh_i^H(\bmR_{\ibar}^{-1}-\bmR_{i}^{-1})\bmHh_i\right)\textrm{vec}(\bmV_k) \\ \implies
\textrm{vec}(\bmVh_k) = \textrm{G.E.V}(\bmS_{t,k},\bmI_{t,k}).
\end{array}
\label{eq_gev}
\eeq
In \eqref{eq_gev}, we take the dominant $D$ eigen vectors as the solution for $\bmV_k$.
Following similar procedure, with $\bmV_k$ fixed, we can derive the combiner matrices as in \eqref{eq_gev_theorem}.\hspace{45mm}$\blacksquare$

Intuitively, \eqref{eq_gev} implies that $\bmV_k$ should be in the orthogonal complement of the semantically aware leakage component. This means that the leakage channel space is weighted by the power in the causal component, and, hence, the null space of $\bmV_k$ contains only those user channels that have semantically rich information. A similar interpretation follows for the combiner matrices, too, with the BF or nulling along the effective channel matrices $\bmHh_k\bmV_k$ or $\bmHh_k\bmV_i$, respectively.
Further, given the BF and combiner matrices, following classical max-min optimization algorithms, we translate the problem as a maximization over an auxiliary variable $\alpha$.
\vspace{-2mm}\beq
\vspace{-1mm}\begin{aligned}
&\mathcal{P}_3:\,\,\, \,\,\,\,\,\,\,\,\,\max\limits_{\bmz}\alpha \\
\,\,\, \,\,\,\,\,\,\,\,\,\,\,\,\,\,\,\,\,\,\,\,\,\mbox{subject\,to}\,\,\,& p\left(E({\bmz}_k,{\widehat{\bmz}}_k) < \delta \right) \geq 1-\epsilon, \\
& \mathbb{E}_{q_k}S_k(\bmz_k;\bmy_k) \geq \alpha, \forall k.
\end{aligned}
\label{eq_P3}
\vspace{-0mm}\eeq
This can solved using the bisection algorithm, as shown in Algorithm~\ref{alg_ta_CSG}. The feasibility \eqref{eq_P3} can be solved using semidefinite programming \cite{BoydCUP2004} after rewriting it $\bmZh_k \!=\! \bmz_k\bmz_k^H$. The resulting instantaneous components of $\bmV_k$, $\bmW_k$, and $\bmz_k$ obtained are sub-optimal (locally optimal solution) due to the approximate convex (originally non-convex) objective functions that get solved. 
\begin{figure*}
\vspace{-4mm}\beq
\vspace{-2mm}\begin{array}{l}
p(\bmz_k^t\mid \bmz_{k}^{t-1})\! = \!\prod\limits_{i\neq I}^d p(\bmz_{k,i}^t\!\mid [\bmM_i\odot\bmz_{k}^{t-1}])^{R_{ki}} \prod\limits_{i\in I}^d p^{\prime}(\bmz_{k,i}^t\mid\! [\bmM_i\odot\bmz_{k}^{t-1}])^{1-R_{ki}}. 
\end{array}\label{eq_ztrans}
\vspace{-1mm}\eeq
\vspace{-2mm}\beq
\vspace{-1mm}\begin{array}{l} 
\!\!p(\bmV_k^t\mid \bmV_{k}^{t-1}\!\!,\mZ^{t-1},\mathcal{H}^{0:t-1}) \!= \!\prod\limits_{i\neq I}^d \!p(\bmV_{k,i}^t\mid [\bmN_i\odot\bmV_{k}^{t-1}],\mZ^{t-1},\mathcal{H}^{0:t-1})^{Q_{ki}} \prod\limits_{i\in I}^d p^{\prime}(\bmV_{k,i}^t\mid [\bmN_i\odot\bmV_{k}^{t-1}],\mZ^{t-1},\mathcal{H}^{0:t-1})^{1-Q_{ki}}.
\end{array}\label{eq_Vk_CGm}
\vspace{-2mm}\eeq\vspace{-2mm}
\end{figure*}
\vspace{-2mm}
\setlength{\textfloatsep}{0pt}
\begin{algorithm}[t]\scriptsize
\caption{Proposed max-min solution}\label{alg_ta_CSG}
 \textbf{Given:} $\bmV_k,\bmW_k$. \textbf{Output:} $\bmzh_k$.\\ 
\textbf{Initialize:} $\alpha^{l} \rightarrow 0$, $\alpha^u \rightarrow $ very large \\ \mbox{Choose larger values for } $\lambda_s, \balpha_s,\balpha_l$. \\
\begin{algorithmic} 
\vspace{-2mm}\STATE \hspace{0.05cm} \textbf{do} {\hspace{0.1cm} (Perform bisection)}
\STATE \hspace{0.1cm} $\alpha^{mid} = \frac{\alpha^l+\alpha^u}{2}$.
\STATE \hspace{0.1cm} \textbf{if} feasible \textbf{then} \\
\STATE \hspace{0.5cm} $\alpha^{l} \leftarrow \alpha^{mid}$ \\
\STATE \hspace{0.5cm} $z_k^{*} \leftarrow z_k \forall k$ based on the solution to the feasibility problem. 
\STATE \hspace{0.1cm} \textbf{else}
\STATE \hspace{0.5cm} $\alpha^{u} \leftarrow \alpha^{mid}$ \\
\end{algorithmic}
\label{algo1}  
\vspace{-0mm}\end{algorithm}

\vspace{-3mm}\subsection{Proposed VCN for BF and Causality Dynamics}
\label{Section_VCN}
\vspace{-3mm}

We have so far looked at computing the portion of the BF, combiner matrices, and the causal representations that depend on the instantaneous channel matrices. Next, we look at computing $\bmVt_k$ and $\bmzt_k$ using the expectation of the posterior distribution of them given the user data, channel and causality history. Since computation of the actual posterior in \eqref{eq_mode_pos} is infeasible, we propose to compute an approximate posterior $q_{\bphi}(\bmz_{k}^t,\bmV_k^t\!\mid\! \mathcal{H}^{0:t-1},\mZ^{0:t-1})$. We assume that $\bmVt_k \!\in\! \mathcal{C}$, where $\mathcal{C}$ represents the codebook (with finite resolution vectors) from which the columns of BF matrices are chosen. For example, the vectors in the codebook here can be columns of DFT matrix of dimensions $M\!\times \!M$. 

For a given dataset of observations (includes multi-modal data and DL channel matrices) and BFs represented as $(\mX^{0:T}, \mY^{0:T},  \mZ^{0:T}, \\ \mV^{0:T})$, we can write the following generative model (superscript denotes dimension time)
\vspace{-1mm}\beq
\vspace{-1mm}
\begin{array}{l}
p(\mX^{0:T}, \mY^{0:T}, \mV^{0:T}) =\int \prod\limits_t\prod\limits_k   p_{\btheta}(\bmx_{k}^t\mid \bmz_{k}^t) p_{\btheta}(\bmz_{k}^t\mid\bmz_{k}^{t-1})\\p_{\btheta}(\bmy_{k}^t\mid \mZ^t,\mV^t,\bmH_k^t)p_{\btheta}(\bmV_{k}^t\mid\bmV_{k}^{t-1},\bmH_{k}^{(0:t-1)})d\bmH_k^td\bmz_{k}^t.
\end{array}
\vspace{1mm}\eeq
We propose to approximate the posterior in \eqref{eq_mode_pos} as $q_{\bphi}(\bmz_{k}^t,\bmV_k^t\mid \mathcal{H}^{0:t-1},\mZ^{0:t-1}) =q_{\bphi}(\bmz_{k}^t \mid \bmZ_k^{0:t-1})q_{\bphi}(\bmV_k^t \mid \mathcal{H}^{0:t-1},\mZ^{0:t-1})$, where $\bphi$ represents the neural network parameters. The approximate posterior can be computed by maximizing the variational evidence lower bound (ELBO) \cite{HoffmanJMLR2013}, which is the principle behind stochastic variational inference and written as
\vspace{-2mm}\beq
\vspace{-2mm}\begin{array}{l}B(\btheta,\bphi) = \sum\limits_{t=0}^T\sum\limits_{k=1}^K\mathbb{E}_{q_{\bphi}}[\log p_{\btheta}(\bmy_{k}^t\mid\mZ^t,\mV^t,\mathcal{H}^{t})]  - \\ \mathbb{E}_{q_{\bphi}}\left[KL\left(q_{\bphi}(\bmz_{k}^t,\bmV_k^t\mid \mathcal{H}^t,\mY^t)|| p(\bmz_{k}^t,\mV^t\mid \bmz_{k}^{t-1},\mV^{t-1})\right)\right].
\end{array}\vspace{1mm}
\eeq
Further, we define a causal structure which we follow for the dynamics of $\bmz_{k}^t$. A causal graphical model (CGM) \cite{PetersMIT2017} is defined as a set of random
variables ${X_1, ..., X_d}$, their joint distribution $P_X$, and a directed acyclic graph (DAG), $G = (X, E)$,
where each edge $(i, j) \in E$ implies that $X_i$
is a direct cause of $X_j$. The joint distribution admits a
causal factorisation such that
$p(X_1,\cdots,X_d) = \prod\limits_{i=1}^d p(X_i\mid \textrm{pa}(X_i))$, where $\textrm{pa}(X_i)$ is the set of parent nodes of $X_i$ in the DAG. We use CGM to define the dynamics of $\bmz_k,\bmV_k$. In contrast to standard graphical models, CGMs propose to include the notion of interventions, i.e., local changes in the causal distribution (representing ``what if $X_i$ was generated in a different way"). Given the set of intervention targets $\mathcal{I}\!\subset\! X$, the changed distribution gets written as
$
p(X_1,\cdots,X_d) = \prod\limits_{i\neq I}^d p(X_i\mid \textrm{pa}(X_i)) \prod\limits_{i\in I}^d p^{\prime}(X_i\mid \textrm{pa}(X_i))$.
The interventions make the proposed algorithm invariant to changes in the dynamics that describe the causal structure. 

Using definition of CGMs, we next look at the dynamics of causal representation $\bmz_k$. We define $\bmM$ as the adjacency matrix with each entry $M_{ij}$ represents whether $z_{k,i}$ causes $z_{k,j}$. We represent the causal transitions across time as
$p(\bmz_{k}^t\mid \bmz_{k}^{t-1}) = \prod_{i}p_i(z_{k,i}^t\mid [\bmM_i\odot \bmz_k^{t-1}]),$
where $\odot$ denotes elementwise product and $\bmM_i$ represents $i^{th}$ colummn of $\bmM$. We use $R_{ki}$ to represent whether $i\!\in\! \mathcal{I}$ for user $k$'s causal state.  The full interventional causal model of the transition
probability can be written as \eqref{eq_ztrans}. This modular adaptation facilitates structural
transfer of knowledge between different wireless environments, and hence, represents an invariant causal representation with respect to source causality distribution. Similarly, we represent the transitions for the BF as in \eqref{eq_Vk_CGm},
where $Q_{ki}$ represents whether $i\!\in\! \mathcal{I}$ and $N_i$ is a matrix with all entries as ones, except columns $i\!\in\! \mathcal{I}$ whose entries are all zeros.

\vspace{-2mm}\vspace{-3mm}\subsection{Proposed Solution}\vspace{-3mm}

VCN \cite{AnsonArxiv2022} learns from
action-observation sequences of an undisturbed environment $( \mX_{(0)}^{0:T},\mY_{(0)}^{0:T},\mV_{(0)}^{0:T},\mathcal{H}_{(0)}^{0:T})$ and $L$ intervened environments $( \mX_{(1:L)}^{0:T},\mY_{(1:L)}^{(0:T)},\mV_{(1:L)}^{(0:T)},\mathcal{H}_{1:L}^{(0:T)})$. The approach of VCN follows the latent state-space model framework, where a latent representation (joint modality) of the multi-modal observations and the channel statistics based component of the BF matrices are learnt with a transition model in \eqref{eq_ztrans} by maximizing the ELBO. 

\vspace{-5mm}\subsection{Training}
\vspace{-3mm}

The task of model training is to determine the model parameters $\btheta$, the approximate posterior
parameters $\bphi$, the causal graph $\mG$, and the intervention targets $\mI$. Each entry $M_{ij}$ of the graph $\mG_k$ follows a Bernoulli distribution with success probability defined as $\sigma(\alpha_{ij})$, where $\sigma$ is the sigmoid function. Similarly, a random binary matrix $\bmR_k$ is parameterized using the scalar $\beta_{ij}$ for each entry. $\balpha, \bbeta$ are the set of all $\alpha_{ij},\beta_{ij}$, respectively. We maximize
the expected ELBO across all environments over causal graphs and intervention masks.
\vspace{-4mm}
\beq
\vspace{-1mm}\begin{array}{l}
\mathcal{L}(\btheta,\bphi,\balpha,\bbeta) = \sum\limits_{l=0}^L \mathbb{E}_{\mathcal{G},\mathcal{I}}B\left(\mX_{(l)}^{0:T},\mY_{(l)}^{0:T};\btheta,\bphi,\mathcal{G},\mathcal{I}\right) \\  -\sum\limits_k\lambda_{\mG_k}\abs{\mG_k} - \sum\limits_k\lambda_{\mI_k}\abs{\mI_k}.
\end{array}
\vspace{-1mm}\eeq
The gradients through the outer expectation and the expectation term in ELBO are estimated using the techniques mentioned in \cite{AnsonArxiv2022}.
\begin{figure}[t]\vspace{-4mm}
\centerline{\includegraphics[width=3.3in,height=1.5in]{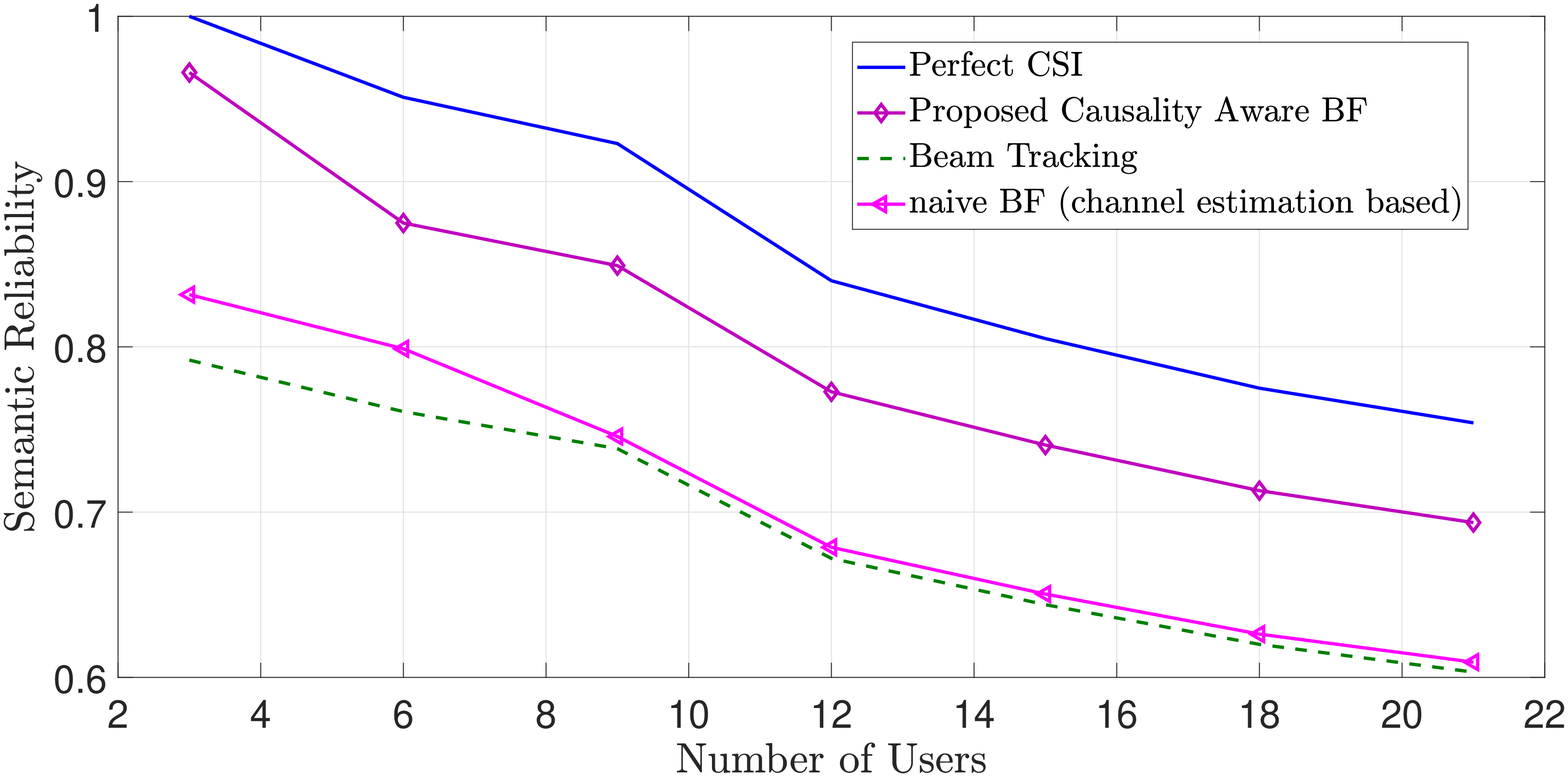}}\vspace{-3.5mm}
\caption{\scriptsize Semantic reliability vs number of users ($M=64$).}
\label{SemanticReliability}\vspace{-0mm}
\vspace{-2mm}
\end{figure}\vspace{-0mm}
\begin{figure}[h]
\vspace{-1mm}
\centerline{\includegraphics[width=3.3in,height=1.5in]{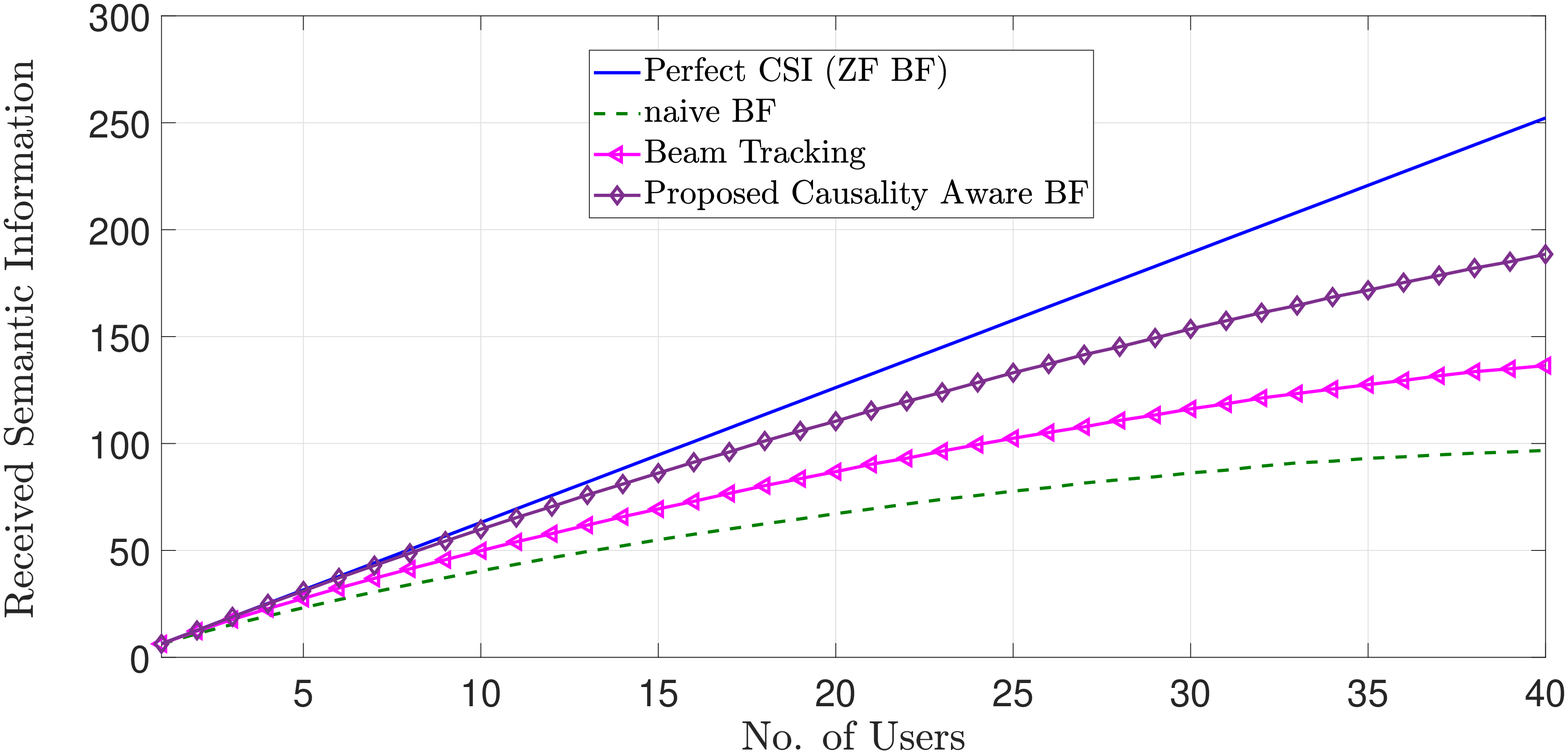}}\vspace{-3.5mm}
\caption{\scriptsize Received semantic information vs number of users ($M=64$).}
\label{SemanticSE_ICASSP}\vspace{-0mm}
\vspace{-3mm}
\end{figure}

\vspace{-4mm}\subsection{ VCN Architecture}
\vspace{-2mm}

The proposed VCN architecture is based on \cite{DavidNIPS2018}. The generation network $p_{\btheta}(\bmx_k^t\mid \bmz_k^t)$ is modeled using variational autoencoder (VAE) \cite{AnsonArxiv2022} and the transition models are represented using the recurrent neural networks (RNN) ]\cite{DavidNIPS2018}.

\vspace{-4mm}\begin{center}
\vspace{-4mm}\begin{table}[t]
\begin{tabular}{|c|c| }
\hline
 Parameters &  Values  \\ 
 \hline
 Operating Frequency & 300 GHz to 450 GHz  \\ 
 $G_t$ & 20 dBi  \\ 
 $G_r$ & 20 dBi  \\ 
 Bandwidth & 1 GHz  \\ 
 Noise Power & -75 dBm \\
 \hline 
\end{tabular}\vspace{-3mm}
 \centering\caption{Simulation Parameters}
\end{table}
\end{center}\vspace{-1mm}

\vspace{-1mm}\section{Simulation Results}
\vspace{-2mm}

We now evaluate the received semantic information at the UEs for the proposed VCN+G.E.V based dynamic BF. To demonstrate the
benefits of our schemes under high mobility, we consider three benchmarks: 1) perfect channel estimate, 2) beam tracking using a DFT codebook \cite{Al-TanDaiJSAC2021}, and 3) BF using least squares estimated channel (naive BF) using UL pilots. ). Existing AI solutions use deep neural networks that comprise classical fully connected layers and nonlinear activation functions, which are not able to generalize to unseen situations during training. In contrast, exploiting a shared causal structure among multiple data distributions leads to better generalization and outperforms conventional schemes for unseen channel environments or transmit data distributions. High mobility refers to a user speed of $90$ km/h here. The channel and observations are generated using \eqref{eq_obs_model} and \eqref{eq_channel_model}. Multi-model causal structure is generated using independent $S=4$ nodes random graph $G$ created from the Erd\"os-R\'enyi (ER) random graph model, similar as \cite{ChristoTWCArxiv2022}.

In Figure.~\ref{SemanticReliability}, we show the semantic reliability versus the number of users in the network. Clearly, our proposed scheme has better semantic reliability (around $14\%$) compared to the conventional methods based on beam tracking using DFT codebooks.

Figure.~\ref{SemanticSE_ICASSP} shows the received semantic information computed using \eqref{eq_semInfo_listener} versus the number of users. The proposed causality aware BF scheme is shown to have superior performance (around $43\%$ improvement) compared to conventional schemes such as beam tracking using DFT codebook and the naive scheme that uses an LMMSE estimate of the channel (using the uplink pilots and does not explicitly model the Doppler effects) to derive the ZF BF.

\vspace{-2mm}
\section{Conclusions}
\vspace{-2mm}

In this paper, we have introduced a novel BF framework for multi-user wireless systems in THz bands. The proposed BF solution is guided by user data causality and channel dynamics and helps to achieve semantically reliable data transmission. Simulation results demonstrate the superior reliability and semantic information for the proposed scheme compared to conventional methods.
\vspace{-2mm}

\newpage
\bibliographystyle{IEEEbib}
\def\baselinestretch{0.9}\selectfont
\bibliography{semantic_refs}

\end{document}